\documentclass[preprint]{aastex}
\usepackage{graphicx}
\usepackage{txfonts}
\usepackage{natbib}
\usepackage{longtable}
\usepackage{rotating}

\bibpunct{(}{)}{;}{a}{}{,}

\begin{document}

\title{A High-Resolution Study of the CO-H$_2$ Conversion Factor in the Diffuse Cloud MBM 40}

\author{David L. Cotten and Loris Magnani}
\affil{Department of Physics and Astronomy, University of Georgia, Athens, GA, 30602}

\begin{abstract}
\singlespace
We made CO(1-0) observations of 103 lines of sight in the core and envelope of the 
high-latitude cloud MBM 40 to determine how the CO-H$_2$ conversion 
factor ($\mathrm{X_{CO}}$) varies throughout the cloud.  Calibrating $\mathrm{X_{CO}}$ with CH data at 
similar resolution (1$\arcmin$ for CO, 1.5$\arcmin$ for CH) yields values of $\mathrm{X_{CO}}$ 
ranging from 0.6 $\times$ 10$^{20}$ to 3.3 $\times$ 10$^{20}$ cm$^{-2}$ 
[K km s$^{-1}$]$^{-1}$ with an average of 1.3  $\times$ 10$^{20}$ cm$^{-2}$ 
[K km s$^{-1}$]$^{-1}$.
Given that the cloud has a peak reddening of 0.24 mag, it should be classed as a diffuse
rather than a translucent molecular cloud.  The mass obtained from the CO data and our
values of $\mathrm{X_{CO}}$ is 9.6 M$_\odot$ for the core, 12 M$_\odot$ for the envelope, and
10 M$_\odot$ for the periphery of the cloud.   A third of the molecular mass of the cloud is found
in a region with E(B-V) $<$ 0.12 mag.   With these mass estimates, we determine
that the cloud is not gravitationally bound.
\end{abstract}

\keywords{ISM:molecules, ISM:clouds, ISM:abundances}


\section{Introduction}\label{sec:intro}

The empirical conversion factor between N(H$_2$) and the velocity-integrated
CO(1-0) main beam antenna temperature ($\int$ T$_{mb}$ dv $\equiv$ $\mathrm{W_{CO}}$) is
routinely used to determine the total molecular content of an interstellar cloud
mapped in CO.  Sometimes known as $\mathrm{X_{CO}}$ ($\equiv$ N(H$_2)$/$\mathrm{W_{CO}}$), in most
Galactic and extragalactic molecular studies the conversion factor
 is taken to be  constant with a
typical value of 1.8 $\times$ 10$^{20}$ cm$^{-2}$ [K km s$^{-1}$]$^{-1}$  (e.g.,
Dame et al. 2001; we will drop the units of $\mathrm{X_{CO}}$ in the remainder of the
paper for brevity).  However, in the class of small molecular clouds with 1 mag
$<$ A$_V$ $<$ 5 mag known as translucents, the value of $\mathrm{X_{CO}}$ has been shown
to vary from cloud-to-cloud and even over a given cloud \citep{mag95,mag98}.  In
calibrating $\mathrm{X_{CO}}$, a surrogate tracer for N(H$_2$) must be employed.
Traditional techniques involve using the diffuse gamma-ray background, the
assumption of virial equilibrium for the cloud in question, and the extinction
produced by dust in the cloud \citep[see review by][]{com91}.  For translucent
clouds these methods are not ideal given the low gas and dust column densities
and the absence
 of virial equilibrium for many of
the clouds \citep{mag95}.  Thus, two more suitable methods were devised: (1)
Using the infrared emission from the dust in a cloud, assuming a standard
gas-to-dust ratio, and correcting for the dust associated with atomic gas along
the line of sight \citep{deV87}; and (2) using the CH $^2\Pi_{1/2}$ (F=1-1)
hyperfine, ground state transition at 3335 MHz to determine N(CH), and then the
linear relationship between N(CH) and N(H$_2$) at low extinction to obtain the
latter quantity (e.g., Magnani \& Onello 1995).

The infrared method requires that an estimate of N(HI) along the given line of
sight be made so that the emission from the dust associated with the atomic gas
can be subtracted from the overall dust emission.  This is somewhat problematic
given that most available HI surveys are at low-resolution (21 -
45$\arcmin$)$\footnote{The GALFA HI survey from Arecibo is at significantly
better resolution, but covers only those regions with $-$2$\arcdeg \le \delta
\le$ 38$\arcdeg$.} $ while CO(1-0) observations are often at $\sim$ 1$\arcmin$
resolution. Another issue involves the assumption that the infrared  emissivity
per hydrogen nucleon is the same for dust mixed with both the atomic and
molecular components \citep[see][for a discussion]{mag95}. Despite these issues,
the infrared method has been used successfully for translucent clouds and often
leads to values for $\mathrm{X_{CO}}$ less than 1 x 10$^{20}$  \citep{deV87}.

Observations of  the CH ground state, hyperfine, main line transition
at 3335 MHz and  the CO(1-0) line at 115 GHz made at similar angular resolution can
be used to determine $\mathrm{X_{CO}}$ as was done by \citet{mag95} for a sample of
translucent and dark clouds.  They determined that for 28 lines of sight in 18
translucent clouds $\mathrm{X_{CO}}$ varied from 0.3 $\times$ 10$^{20}$ to 6.8 $\times$
10$^{20}$.  Although the CH 3335 MHz line is not a good tracer of H$_2$ in
high-extinction, high column density regions, Magnani \& Onello applied the
technique to 13 lines of sight in 5 dark clouds and obtained $\mathrm{X_{CO}}$ values
ranging from 0.8 $\times$ 10$^{20}$ to 8.6 $\times$ 10$^{20}$.  In a subsequent
study, the CH method was applied to different regions of 2 translucent clouds
and significant variation across the face of these clouds was seen: In MBM 16,
$\mathrm{X_{CO}}$ varied from 1.6 $\times$ 10$^{20}$ to 17.3 $\times$ 10$^{20}$, and in
MBM 40 from 0.7 $\times$ 10$^{20}$ to 9.7 $\times$ 10$^{20}$.  These studies
were made at resolutions of 8-9$\arcmin$. \citet{ccm10} studied CH emission in
two translucent clouds at significantly higher resolution (1.5$\arcmin$ for the 
CH 3335 MHz line
and 45$\arcsec$ for CO(1-0)) and determined that $\mathrm{W_{CO}}$ and the
velocity-integrated CH line strength at 3.3 GHz ($\int$ T$_B$ dv $\equiv$
$\mathrm{W_{CH}}$, where T$_B$ is the brightness temperature) did not correlate well.  
Unfortunately, \citet{ccm10} did not
calculate $\mathrm{X_{CO}}$ for the two clouds they observed, so one of the goals of this
paper is to determine $\mathrm{X_{CO}}$ for one of those clouds (MBM 40) that has
sufficient data for statistical analysis.

The lack of correlation between $\mathrm{W_{CO}}$ and $\mathrm{W_{CH}}$ in MBM 16 and 40
implies that either CO or CH, or possibly both, do
not correlate well with H$_2$ in lower extinction clouds.  If CH is assumed to
correlate fairly well with H$_2$ in low extinction regions - as all other
studies to date have indicated (see references in \S 1), then the lack
of correlation between $\mathrm{W_{CO}}$ and $\mathrm{W_{CH}}$ could be interpreted as a variation
in $\mathrm{X_{CO}}$. \citet{lp12} found significant variability for  $\mathrm{X_{CO}}$ on
arcminute scales in diffuse molecular clouds and attribute this variability
primarily to radiative transfer and CO chemistry effects. 

Like the infrared method, the CH method also has difficulties associated with it.  
To derive N(CH), it is
normally assumed that $\vert T_{ex}(3335) \vert \gg$ T$_{BG}$; however, some
studies of low-extinction clouds imply otherwise \citep{lein84,jm85}.  Most
researchers just assume (as we will - but see \S 3) that the excitation temperature of the 3335
MHz line is far from the value of T$_{BG}$, but optical and UV absorption observations as described
by \citet{lein84} are needed to really establish the value of T$_{ex}$(3335) for 
a given line of sight. Also, while one
can certainly ascribe the lack of correlation between $\mathrm{W_{CO}}$ and $\mathrm{W_{CH}}$ to
variations in $\mathrm{X_{CO}}$, they can also be due to variations in the CH abundance
\citep{ll02}.  However, a recent study by \citet{lp12} seems to show
definitively that in low-extinction clouds the variation in CO excitation and
abundance is significant both from cloud-to-cloud and even over different
regions of a given cloud. In this paper, we will assume that, at low extinction,
$\mathrm{W_{CH}}$ can lead to robust estimates of N(CH) and that this quantity is a
linear tracer of N(H$_2$).

\begin{figure}
\includegraphics[width=\textwidth]{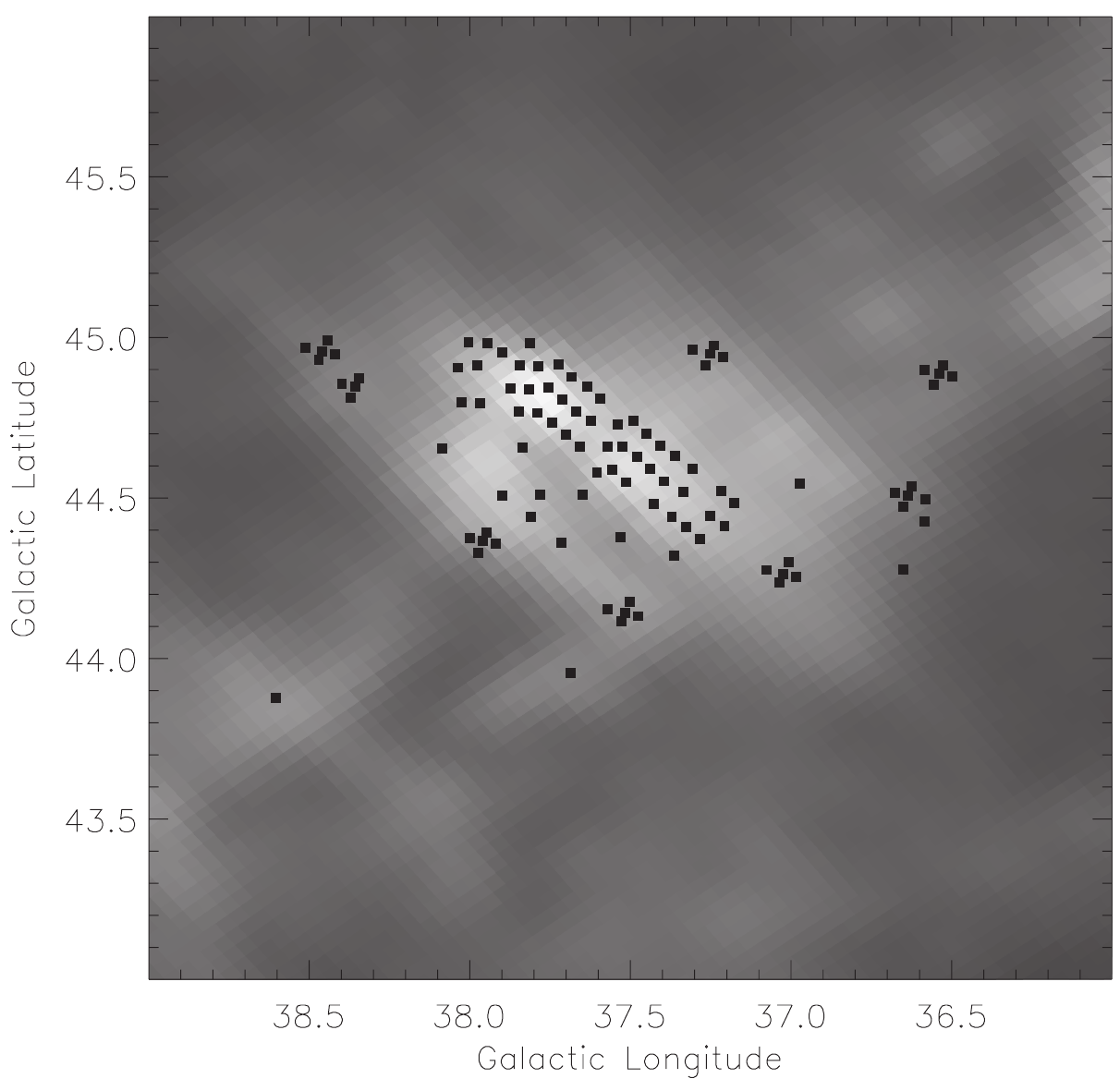}
\caption{Positions of CO(1-0) observations for MBM 40 superimposed on an E(B-V) 
map from the \citet{SFD98} data. Coordinates are Galactic $\ell$ and ${\it b}$ 
centered on ($\ell$, ${\it b}$) = 37.4$^{\circ}$ and 44.5$^{\circ}$, respectively. The
black squares represent the positions of our CO(1-0) observations. The coordinates and results of these 
observations are listed in Table~\ref{tab:WCO} with averages for the detections in
each region in Table~\ref{tab:avgco}. The dust emission from the cloud is in units of E(B-V) magnitudes on a
linear scale ranging from 0.05 to 0.24 mag.}\label{fig:COpos}
\end{figure}

The paper is organized as follows: In \S 2 we describe the new CO observations
of MBM 40 that were made at selected regions in the core and envelope of the
cloud.  In \S3 we will use the CH method to calibrate $\mathrm{X_{CO}}$ at high
resolution ($\sim$ 1.5$\arcmin$) for the cloud MBM 40 (previously classified as a translucent  cloud), where the CH data
comes from \citet{ccm10}.
From our CO observations and the calibrated $\mathrm{X_{CO}}$ values we will determine
 the mass
in the core, envelope, and periphery of the cloud (\S 4) and discuss the
gravitational stability of this object. Finally, in \S5 we summarize our results.


\section{Observations}

Observations of the CO(1-0) rotational transition were made at the Arizona Radio
Observatory (ARO) 12 m telescope at Kitt Peak National Observatory in December
of 2008$\footnote{The 12 m is part of the Arizona Radio Observatory and is
operated by the University of Arizona with additional funding by the Mt. Cuba
Astronomical Foundation.}$. The angular resolution of the telescope at 115 GHz
was 1$\arcmin$ and the observations were made in position-switching mode with
the off-source taken to be one degree east or west of the target in azimuth. The
off positions were checked to be relatively free of dust emission using the
Schlegel, Finkbeiner, \& Davis (1998; hereafter SFD) dust maps. The spectrometer consisted of 100 kHz
and 250 kHz filterbanks that  provide velocity coverages of 62 and 167 km
$\mathrm{s^{-1}}$, respectively, with velocity resolutions of 0.26 and 0.65 km
$\mathrm{s^{-1}}$, respectively.

\begin{deluxetable}{ c  c  c c  c   c   c   c }
\tabletypesize{\small}
\tablecaption{Observations of the CO(1-0) line in MBM 40.\label{tab:WCO}}
\tablehead{
\colhead{RA (2000)} & \colhead{Dec (2000)} & \colhead{$\ell$} & \colhead{\it b} & \colhead{$\mathrm{T_{R}^{*}}$} &
\colhead{$\mathrm{\Delta v(FWHM)}$} & \colhead{$\mathrm{v_{LSR}}$} & \colhead{W$\mathrm{_{CO}}\tablenotemark{a}$}   \\
\colhead{(h m s)} & \colhead{($^\circ$ ' ")} & \colhead{deg} & \colhead{deg} &  \colhead{(K)} &
\colhead{(km s$^{-1}$)} & \colhead{(km s$^{-1}$)} & \colhead{(K km s$^{-1}$)}
}
\startdata
\multicolumn{6}{c}{$\bold\underline{Periphery\ Region}$}\\
16 08 36.0	&	21 11 24	& 36.527	& 44.914	& 0.17	+/-	0.05	&	1.01	+/-	0.33	&	 4.50	&	0.21	+/-	0.07 \\
16 08 43.2	&	21 09 36	& 36.498	& 44.878	& 0.39	+/-	0.07	&	0.80	+/-	0.14	&	 2.95	&	0.40	+/-	0.07 \\
16 08 43.2	&	21 11 24	& 36.539	& 44.887	&		0.070 $\tablenotemark{b}$	& 		& 		& 			\\					
16 08 43.2	&	21 13 48	& 36.585	& 44.898	&		0.070		& 				& 		& 			 	\\					
16 08 52.8	&	21 11 24	& 36.555	& 44.852	&		0.070		& 				& 		& 			 	\\					
16 10 00.0	&	22 30 36	& 38.444	& 44.991	& 0.27	+/-	0.07	&	0.58	+/-	0.16	&	 3.63	&	0.20	+/-	0.05 \\
16 10 09.6	&	22 28 48	& 38.419	& 44.948	& 2.29	+/-	0.07	&	0.64	+/-	0.02	&	 3.70	&	1.88	+/-	0.06 \\
16 10 09.6	&	22 30 36	& 38.459	& 44.956	& 0.25	+/-	0.08	&	0.81	+/-	0.25	&	 3.68	&	0.26	+/-	0.08 \\
16 10 09.6	&	22 33 00	& 38.513	& 44.967	&		0.070		&				&		& 			   	\\ 				
16 10 14.4	&	21 08 24	& 36.624	& 44.535	&		0.070		& 				& 			& 			\\				
16 10 16.8	&	22 30 36	& 38.470	& 44.929	& 0.42	+/-	0.08	&	0.66	+/-	0.12	&	 4.03	&	0.35	+/-	0.06 \\
16 10 21.6	&	21 06 00  & 36.582	& 44.496	& 0.55	+/-	0.07	&	0.52	+/-	0.07	&	 2.99	&	0.36	+/-	0.05 \\
16 10 21.6	&	21 08 24	& 36.636	& 44.508	&		0.060	&				&	 		&		 		\\
16 10 21.6	&	21 10 12	& 36.676	& 44.517	&		0.060	&				&			&		 		\\					
16 10 24.0	&	22 24 36	& 38.346	& 44.874	& 1.85	+/-	0.06	&	0.76	+/-	0.03	&	 3.43	&	1.80	+/-	0.06 \\
16 10 31.2	&	21 08 24	& 36.651	& 44.472	& 0.11	+/-	0.06	&	0.92	+/-	0.50	&	 4.35	&	0.13	+/-	0.07 \\
16 10 31.2	&	22 24 36	& 38.357	& 44.848	& 1.37	+/-	0.07	&	0.68	+/-	0.04	&	 3.49	&	1.19	+/-	0.06 \\
16 10 31.2	&	22 26 24	& 38.398	& 44.856	& 0.93	+/-	0.07	&	0.63	+/-	0.05	&	 3.52	&	0.74	+/-	0.05 \\
16 10 40.8	&	22 24 36	& 38.372	& 44.812	& 0.88	+/-	0.09	&	0.71	+/-	0.07	&	 3.49	&	0.80	+/-	0.08 \\
16 12 19.2	&	22 00 00 	& 37.975	& 44.330	&		0.104	&				&				&			\\					
16 13 36.0	&	21 41 24	& 37.685	& 43.955	& 0.61	+/-	0.05	&	1.10	+/-	0.09	&	 1.89	&	0.85	+/-	0.07 \\
\multicolumn{6}{c}{$\bold\underline{Envelope\ Region}$}\\
16 06 19.2	&	20 52 12	& 35.864	& 45.323	& 2.90	+/-	0.11	&	0.51	+/-	0.02	&	 3.49	&	1.88	+/-	0.07 \\
16 06 19.2	&	20 54 36	& 35.918	& 45.335	& 2.84	+/-	0.11	&	0.60	+/-	0.02	&	 3.45	&	2.17	+/-	0.09 \\
16 09 00.0	&	21 41 24	& 37.240	& 44.975	& 0.19	+/-	0.06	&	0.68	+/-	0.22	&	 3.57	&	0.16	+/-	0.05 \\
16 09 07.2	&	21 39 36	& 37.211 	& 44.940	&		0.068	&				& 			& 	 			\\				
16 09 07.2	&	21 41 24	& 37.252	& 44.949	&		0.063	& 				&			& 			 	\\					
16 09 07.2	&	21 43 48	& 37.306	& 44.961	& 0.31	+/-	0.07	&	1.20	+/-	0.27	&	 3.25	&	0.48	+/-	0.11 \\
16 09 16.8	&	21 41 24	& 37.267	& 44.913	& 0.27	+/-	0.07	&	0.34	+/-	0.09	&	 4.17	&	0.12	+/-	0.03 \\
16 09 28.8	&	22 04 48	& 37.813	& 44.983	& 0.71	+/-	0.08	&	1.00	+/-	0.12	&	 2.72	&	0.91	+/-	0.11 \\
16 09 36.0	&	22 10 12	& 37.946	& 44.983	& 1.47	+/-	0.08	&	0.82	+/-	0.05	&	 2.65	&	1.54	+/-	0.09 \\
16 09 38.4	&	22 12 36	& 38.004	& 44.985	& 1.54	+/-	0.1	&	0.84	+/-	0.05	&	 2.83	&	1.65	+/-	0.10 \\
16 10 00.0	&	22 12 36	& 38.038	& 44.906	& 5.22	+/-	0.11	&	0.95	+/-	0.02	&	 3.00	&	6.30	+/-	0.13 \\
16 10 26.4	&	22 10 12	& 38.026	& 44.797	& 6.70	+/-	0.1	&	0.68	+/-	0.01	&	 2.97	&	5.81	+/-	0.09 \\
16 10 31.2	&	21 22 48	& 36.972	& 44.545	&		0.075	&				&			&			 	\\					
16 10 50.4	&	21 32 24	& 37.218	& 44.522	& 0.82	+/-	0.09	&	0.53	+/-	0.06	&	 3.07	&	0.56	+/-	0.06 \\
16 10 57.6	&	21 30 00	& 37.176	& 44.484	&		0.096	&				& 			&	 		 	\\
16 11 04.8	&	22 10 12	& 38.086	& 44.655	& 2.49	+/-	0.09	&	0.89	+/-	0.03	&	 3.37	&	2.80	+/-	0.10 \\
16 11 16.8	&	21 30 00	& 37.207	& 44.413	&		0.106	&				& 			&	 			\\
16 11 16.8	&	21 49 48	& 37.649	& 44.511	& 1.49	+/-	0.08	&	0.72	+/-	0.04	&	 3.32	&	1.37	+/-	0.07 \\
16 11 24.0	&	21 55 12	& 37.781	& 44.511	& 2.19	+/-	0.08	&	0.85	+/-	0.03	&	 3.04	&	2.36	+/-	0.08 \\
16 11 31.2	&	21 32 24	& 37.284	& 44.372	& 1.12	+/-	0.09	&	0.99	+/-	0.08	&	 3.39	&	1.41	+/-	0.11 \\
16 11 33.6	&	21 19 48	& 37.008	& 44.299	& 0.41	+/-	0.09	&	0.38	+/-	0.08	&	 3.07	&	0.20	+/-	0.04 \\
16 11 43.2	&	21 18 00	& 36.984	& 44.255	& 		0.090		&	 		 	& 		&				\\			
16 11 43.2	&	21 19 48	& 37.024	& 44.264	& 0.68	+/-	0.09	&	0.43	+/-	0.06	&	 3.00	&	0.37	+/-	0.05 \\
16 11 43.2	&	21 22 12	& 37.077	& 44.276	&		0.088	&				&			&		 		\\					
16 11 43.2	&	21 42 36	& 37.530	& 44.378	& 0.99	+/-	0.10	&	1.02	+/-	0.10	&	 3.32	&	1.28	+/-	0.13 \\
16 11 48.0	&	21 34 48	& 37.365	& 44.321	& 1.93	+/-	0.13	&	0.81	+/-	0.05	&	 3.19	&	1.99	+/-	0.13 \\
16 11 50.4	&	21 19 48	& 37.036	& 44.237	&		0.081	& 				& 			& 		 		\\			
16 11 57.6	&	21 49 48	& 37.714	& 44.360	& 1.35	+/-	0.08	&	0.79	+/-	0.05	&	 3.16	&	1.36	+/-	0.08 \\
16 12 02.4	&	22 00 00	& 37.949	& 44.393	& 1.09	+/-	0.11	&	0.54	+/-	0.05	&	 3.32	&	0.74	+/-	0.07 \\
16 12 09.6	&	21 58 12	& 37.920	& 44.357	& 0.72	+/-	0.10	&	0.82	+/-	0.11	&	 3.00	&	0.75	+/-	0.10 \\
16 12 09.6	&	22 00 00	& 37.960	& 44.366	& 0.64	+/-	0.09	&	0.58	+/-	0.08	&	 3.06	&	0.47	+/-	0.06 \\
16 12 09.6	&	22 01 48	& 38.000	& 44.375	& 0.50	+/-	0.10	&	0.88	+/-	0.18	&	 3.08	&	0.56	+/-	0.12 \\
16 12 31.2	&	21 37 48	& 37.501	& 44.177	& 0.43	+/-	0.07	&	0.25	+/-	0.04	&	 3.13	&	0.14	+/-	0.02 \\
16 12 40.8	&	21 36 00	& 37.476	& 44.132	& 0.33	+/-	0.11	&	0.94	+/-	0.30	&	 3.18	&	0.39	+/-	0.13 \\
16 12 40.8	&	21 37 48	& 37.516	& 44.141	& 0.37	+/-	0.10	&	0.45	+/-	0.13	&	 3.22	&	0.21	+/-	0.06 \\
16 12 40.8	&	21 40 12	& 37.570	& 44.153	& 0.46	+/-	0.11	&	0.42	+/-	0.10	&	 3.21	&	0.24	+/-	0.06 \\
16 12 48.0	&	21 37 48	& 37.528	& 44.115	&		0.104	& 				& 			& 				 \\				
16 14 45.6	&	22 18 00	& 38.604	& 43.877	&		0.100	&				&			&				 \\
\multicolumn{6}{c}{$\bold\underline{Core\ Region}$}\\
16 09 40.8	&	22 00 00	& 37.724 	& 44.916	& 2.27	+/-	0.07	&	1.12	+/-	0.03	&	 2.86	&	3.24	+/-	0.10 \\
16 09 40.8	&	22 07 48	& 37.900	& 44.953	& 3.92	+/-	0.09	&	0.70	+/-	0.02	&	 2.84	&	3.47	+/-	0.08 \\
16 09 45.6	&	22 02 24	& 37.786	& 44.910	& 4.51	+/-	0.07	&	0.81	+/-	0.01	&	 2.84	&	4.66	+/-	0.08 \\
16 09 48.0	&	21 57 36	& 37.682	& 44.877	& 2.58	+/-	0.10	&	0.69	+/-	0.03	&	 3.34	&	2.27	+/-	0.09 \\
16 09 48.0	&	22 04 48	& 37.844	& 44.912	& 6.47	+/-	0.09	&	0.93	+/-	0.01	&	 3.00	&	7.67	+/-	0.10 \\
16 09 52.8	&	21 55 12	& 37.635	& 44.848	& 2.00	+/-	0.14	&	0.72	+/-	0.05	&	 3.56	&	1.84	+/-	0.13 \\
16 09 55.2	&	22 10 12	& 37.977	& 44.912	& 3.82	+/-	0.08	&	0.92	+/-	0.02	&	 2.85	&	4.49	+/-	0.10 \\
16 10 00.0	&	21 52 48	& 37.593	& 44.810	& 1.67	+/-	0.10	&	0.71	+/-	0.04	&	 3.54	&	1.51	+/-	0.09 \\
16 10 00.0	&	22 00 00	& 37.755	& 44.845	& 5.14	+/-	0.07	&	1.14	+/-	0.02	&	 3.02	&	7.44	+/-	0.11 \\
16 10 02.4	&	22 07 48	& 37.968	& 44.794	& 5.33	+/-	0.08	&	1.00	+/-	0.02	&	 2.82	&	6.77	+/-	0.11 \\
16 10 04.8	&	22 02 24	& 37.816	& 44.839	& 6.24	+/-	0.09	&	1.13	+/-	0.02	&	 2.92	&	9.01	+/-	0.12 \\
16 10 07.2	&	21 57 36	& 37.712	& 44.806	& 6.32	+/-	0.08	&	1.03	+/-	0.01	&	 3.10	&	8.32	+/-	0.11 \\
16 10 07.2	&	22 04 48	& 37.874	& 44.841	& 5.31	+/-	0.08	&	1.01	+/-	0.02	&	 2.77	&	6.85	+/-	0.10 \\
16 10 12.0	&	21 47 24	& 37.491	& 44.739	& 0.91	+/-	0.11	&	1.56	+/-	0.19	&	 3.55	&	1.80	+/-	0.22 \\
16 10 14.4	&	21 55 12	& 37.670	& 44.768	& 7.21	+/-	0.15	&	0.85	+/-	0.02	&	 3.24	&	7.81	+/-	0.16 \\
16 10 16.8	&	21 49 12	& 37.539	& 44.730	& 2.04	+/-	0.09	&	0.86	+/-	0.04	&	 3.46	&	2.25	+/-	0.10 \\
16 10 19.2	&	21 45 00	& 37.449	& 44.700	& 1.33	+/-	0.09	&	1.09	+/-	0.08	&	 3.49	&	1.86	+/-	0.13 \\
16 10 19.2	&	21 52 48	& 37.624	& 44.739	& 5.92	+/-	0.16	&	0.76	+/-	0.02	&	 3.20	&	5.73	+/-	0.16 \\
16 10 21.6	&	22 00 00 	& 37.789	& 44.765	& 5.90	+/-	0.08	&	1.04	+/-	0.01	&	 2.99	&	7.81	+/-	0.10 \\
16 10 24.0	&	22 02 24	& 37.847	& 44.768	& 6.06	+/-	0.09	&	0.80	+/-	0.01	&	 2.85	&	6.19	+/-	0.09 \\
16 10 26.4	&	21 42 36	& 37.407	& 44.662	& 1.49	+/-	0.09	&	1.03	+/-	0.06	&	 3.60	&	1.96	+/-	0.12 \\
16 10 26.4	&	21 57 36	& 37.743	& 44.735	& 7.77	+/-	0.08	&	0.66	+/-	0.01	&	 3.11	&	6.50	+/-	0.07 \\
16 10 31.2	&	21 40 12	& 37.361	& 44.632	& 1.70	+/-	0.10	&	1.20	+/-	0.07	&	 3.51	&	2.61	+/-	0.16 \\
16 10 33.6	&	21 47 24	& 37.526	& 44.659	& 7.06	+/-	0.10	&	0.96	+/-	0.01	&	 3.33	&	8.62	+/-	0.12 \\
16 10 33.6	&	21 55 12	& 37.700	& 44.697	& 6.80	+/-	0.13	&	0.66	+/-	0.01	&	 3.11	&	5.74	+/-	0.11 \\
16 10 36.0	&	21 49 12	& 37.570	& 44.659	& 5.62	+/-	0.08	&	0.90	+/-	0.01	&	 3.48	&	6.47	+/-	0.10 \\
16 10 38.4	&	21 37 12	& 37.305	& 44.591	& 0.84	+/-	0.08	&	0.89	+/-	0.08	&	 3.46	&	0.96	+/-	0.09 \\
16 10 38.4	&	21 45 00	& 37.480	& 44.629	& 5.76	+/-	0.10	&	0.97	+/-	0.02	&	 3.40	&	7.14	+/-	0.13 \\
16 10 38.4	&	22 04 48	& 36.583	& 44.427	& 6.67	+/-	0.08	&	0.70	+/-	0.01	&	 2.84	&	5.99	+/-	0.07 \\
16 10 40.8	&	21 52 48	& 37.658	& 44.659	& 8.10	+/-	0.14	&	0.59	+/-	0.01	&	 3.15	&	6.05	+/-	0.10 \\
16 10 45.6	&	21 42 36	& 37.438	& 44.591	& 6.40	+/-	0.10	&	0.76	+/-	0.01	&	 3.32	&	6.17	+/-	0.10 \\
16 10 50.4	&	22 00 00	& 37.835	& 44.658	& 1.85	+/-	0.08	&	0.83	+/-	0.04	&	 3.12	&	1.95	+/-	0.09 \\
16 10 52.8	&	21 40 12	& 37.396	& 44.552	& 3.52	+/-	0.11	&	1.00	+/-	0.03	&	 3.41	&	4.48	+/-	0.14 \\
16 10 52.8	&	21 47 24	& 37.556	& 44.588	& 9.11	+/-	0.08	&	0.81	+/-	0.01	&	 3.32	&	9.39	+/-	0.09 \\
16 10 57.6	&	21 37 12	& 37.336	& 44.520	& 2.30	+/-	0.11	&	0.91	+/-	0.04	&	 3.45	&	2.67	+/-	0.13 \\
16 10 57.6	&	21 49 12	& 37.604	& 44.579	& 5.85	+/-	0.10	&	0.81	+/-	0.01	&	 3.41	&	6.00	+/-	0.10 \\
16 11 00.0	&	21 45 00	& 37.514	& 44.549	& 9.84	+/-	0.09	&	0.82	+/-	0.01	&	 3.19	&   10.26 +/-  0.09 \\
16 11 12.0	&	21 32 24 	& 37.253	& 44.443	& 0.96	+/-	0.10	&	1.45	+/-	0.15	&	 3.43	&	1.76	+/-	0.18 \\
16 11 12.0	&	21 40 12	& 37.427	& 44.481	& 7.27	+/-	0.10	&	0.84	+/-	0.01	&	 3.23	&	7.74	+/-	0.11 \\
16 11 19.2	&	21 37 12	& 37.371	& 44.440	& 3.85	+/-	0.12	&	0.83	+/-	0.03	&	 3.28	&	4.07	+/-	0.13 \\
16 11 19.2	&	22 04 48	& 36.651	& 44.277	& 3.68	+/-	0.09	&	0.78	+/-	0.02	&	 3.23	&	3.65	+/-	0.09 \\
16 11 24.0	&	21 34 48	& 37.326	& 44.410	& 2.06	+/-	0.11	&	0.81	+/-	0.04	&	 3.29	&	2.13	+/-	0.11 \\
16 11 31.2	&	22 00 00	& 37.899	& 44.508	& 4.46	+/-	0.08	&	0.83	+/-	0.02	&	 3.10	&	4.71	+/-	0.09 \\
16 11 43.2	&	21 55 12	& 37.811	& 44.440	& 3.16	+/-	0.10	&	0.73	+/-	0.02	&	 3.01	&	2.95	+/-	0.10 \\
\enddata
\tablenotetext{a}{$\mathrm{W_{CO}}$ is $\mathrm{\int T_{mb}\ dv}$ .}
\tablenotetext{b}{If only one number is tabulated, then that is the 1-$\sigma$ rms value.}
\end{deluxetable} 
\clearpage

At the 12 m telescope, the CO(1-0) line antenna temperature,
$\mathrm{T_{A}^{*}}$, is corrected for the spillover and scattering efficiency
of the antenna, resulting in the quantity $\mathrm{T_{R}^{*}}$, the radiation
temperature uncorrected for the antenna-beam coupling efficiency,
$\mathrm{\eta_{mb}}$.  Thus, the main beam antenna temperature,  $\mathrm{T_{mb}}$, is equal to
$\mathrm{T_{R}^{*}}$/$\mathrm{\eta_{mb}}$. For the 12 m telescope at 115 GHz, $\mathrm{\eta_{mb}}$ is
approximately 0.85$\footnote{ARO 12 Users Manual:
$\mathrm{http://aro.as.arizona.edu/12\_obs\_manual/12m\_user\_manual.html}$.}$.
 Another correction factor is the beam
dilution that we assume to be equal to 1 (in other words, the source fills the
beam).  Integration times were chosen to give rms noise values of at worst 0.1 K
in the 100 kHz filterbanks.

A total of 103 lines of sight were observed in the core and envelope region of
MBM 40 with 85 detections (see Figure \ref{fig:COpos} and Table \ref{tab:WCO}).
In columns 1 - 4 of Table~\ref{tab:WCO} we list the positions of the observed
lines of sight, in Right Ascension, Declination, and Galactic coordinates. The antenna
temperature $\mathrm{T_{R}^{*}}$ is listed in column 5, the line width
($\mathrm{\Delta v}$) as the Full Width at Half Maximum (FWHM)
 in column 6, the LSR velocity ($\mathrm{v_{LSR}}$) in
column 7, and column 8 is the integrated CO(1-0) main beam temperature, defined
as $\mathrm{W_{CO}}$.  If no line was detected for a given position we list the
1-$\sigma$ rms level in column 5. We divided MBM 40 into three regions based on
E(B-V) in order to understand the effectiveness of the CO(1-0) line as a
molecular tracer in different extinction regimes. The three regions are defined
as: the ``core'' region where E(B-V) $>$ 0.17, the ``envelope'' region where
0.12 $\le$ E(B-V) $\le$ 0.17, and the ``periphery'' region where E(B-V)  $<$
0.12 mag.  The core region includes the two ridges of strong CO(1-0) emission 
described by \citet{shore03} as comprising a wishbone or hairpin structure.
Of the roughly 2 square degrees that comprise MBM 40 on the SFD dust
maps, the periphery,
envelope, and core regions cover 1.2, 0.61, and 0.20 square degrees,
respectively. In the core region there were 44 CO(1-0) observations all
resulting in detections. The envelope and periphery regions contained 28 out of 38
and 13 out of 21 detections, respectively.  Average line values for the detections
in the core, envelope, and periphery regions are shown in Table \ref{tab:avgco} along
with the averages for all the detections.


\section{Determination of $\mathrm{X_{CO}}$ in MBM 40}

Table \ref{tab:WCO} shows the positions observed in CO along with the $\mathrm{W_{CO}}$
value for that line of sight.   To calibrate $\mathrm{X_{CO}}$, we require observations
of the CH 3335 MHz line at a similar resolution.  These observational data are
tabulated by \citet{ccm10} and used to derive N(CH), the values of which are
reproduced here in Table \ref{tab:xco}.  In the core of the cloud, as defined in
\S 2, there  are 44 lines of sight with CO observations in Table \ref{tab:WCO};
for 32 of these CO data points we have CH data. In the envelope, 4 of the 28 CO
detections  have corresponding CH data.  Unfortunately, no CH observations were
made in the periphery region.

A linear relationship between N(CH) and N(H$_2$) was established more than 3 decades ago
and has been repeatedly confirmed \citep{sj82, fed82, dfl84, mat86, mag95, ll02, sheff08, wes10}.  Based on these works, we will assume that
the CH/H$_2$ ratio in diffuse and translucent molecular clouds is 4 x 10$^{-8}$,
good to about 20\%.  This allows us to obtain N(H$_2$) once N(CH) is determined,
but to derive N(CH) from observations of the 3335 MHz line it is
necessary to assume that $\vert T_{ex}$ $\vert$
  $\gg$ T$_{BG}$ for that transition, which may not necessarily be the case. 
On the one hand, \citet{lein84} explicitly measured T$_{ex}$ for three diffuse lines of sight and 
obtained values between $-$1 and 1 K.  On the other, \citet{gen79} found T$_{ex}$
= $-$60 $\pm$ 30 in an on-source/off-source radio observation of a 
a background quasar through the dark cloud L1500.  What is
the situation for MBM 40?  Although direct measurements of T$_{ex}$ for
the 3335 MHz transition do not exist,
we can argue indirectly that if W$_{CH}$ is proportional to N(H$_2$), then the most
likely explanation is that $\vert T_{ex}$ $\vert$  $\gg$ T$_{BG}$.  \citet{ccm10}
show that W$_{CH}$ is proportional to W$_{CO}$ in MBM 40, despite a small
positional offset between the two tracers (likely due to some portion of the CH 3335 MHz line tracing molecular gas
not detectable by the CO(1-0) line).  While the CO(1-0) line may not trace all the
molecular gas in a cloud, it certainly traces most of it \nocite{dht01, RJC05}, (e.g., Dame, Hartmann,
and Thaddeus (2001) in general, and Chastain (2008) in particular for MBM 40).   Thus,  W$_{CH}$ is proportional
to N(H$_2$) in MBM 40.  Since N(CH) is also proportional to N(H$_2$), we
can assume that, for MBM 40, W$_{CH}$ $\propto$ N(CH) and 
so $\vert T_{ex}$ $\vert$  $\gg$ T$_{BG}$.  

With this assumption, N(H$_2$) is readily derived and 
dividing N(H$_2$) by $\mathrm{W_{CO}}$ immediately yields  $\mathrm {X_{CO}}$.
 The resulting values are  tabulated in Table \ref{tab:xco}; they  range from 0.6 to 3.3 $\times$  $\mathrm{10^{20}}$  with an average
 value of 1.3 x $\mathrm{10^{20}}$.   This compares
 reasonably with the values of $\mathrm{X_{CO}}$ obtained by \citet{mag98} at nearly an
 order of magnitude worse resolution; they obtained an average value of 2.6
 $\times$ 10$^{20}$ for the core region only of MBM 40.  

\begin{deluxetable}{ c c c c c c}
\tablecaption{Determination of  $\mathrm {X_{CO}}$ in MBM 40.\label{tab:xco}}
\tablehead{
\colhead{RA (2000)} & \colhead{Dec (2000)} &\colhead{$\ell$} & \colhead{\it b} & \colhead{$\mathrm{N(CH)\ ^{a}}$} & 
\colhead{$\mathrm {X_{CO}}$}\\
\colhead{(h m s)} & \colhead{($^\circ$ ' ")} & \colhead{deg} & \colhead{deg} &  \colhead{$\mathrm{x\ 10^{13}\ (cm^{-2})}$ } &   \colhead{$\mathrm{x\ 10^{20}\ (cm^{-2}\ K\  km\ s^{-1}})$}
}
\startdata
16 09 40.8	&	22 07 48	& 37.900	& 44.953	&	2.7	+/-	0.58	&	1.94	+/-	0.29\\	
16 09 45.6	&	22 02 24	& 37.786	& 44.910	&	3.2	+/-	0.73	&	1.72	+/-	0.30\\	
16 09 48.0	&	22 04 48	& 37.844	& 44.912	&	2.8	+/-	0.47	&	0.91	+/-	0.26\\	
16 09 55.2	&	22 10 12	& 37.977	& 44.912	&	2.7	+/-	0.70	&	1.50	+/-	0.33\\	
16 10 00.0	&	22 00 00	& 37.755	& 44.845	&	3.4	+/-	0.48	&	1.14	+/-	0.25\\	
16 10 00.0	&	22 12 36	& 38.038	& 44.906	&	1.7	+/-	0.43	&	0.67	+/-	0.32\\	
16 10 02.4	&	22 07 48	& 37.968	& 44.794	&	1.6	+/-	0.47	&	0.59	+/-	0.36	\\
16 10 04.8	&	22 02 24	& 37.816	& 44.839	&	2.9	+/-	0.48	&	0.80	+/-	0.26\\	
16 10 07.2	&	21 57 36	& 37.712	& 44.806	&	4.2	+/-	0.58	&	1.26	+/-	0.24	\\
16 10 07.2	&	22 04 48	& 37.874	& 44.841	&	1.7	+/-	0.44	&	0.62	+/-	0.33\\	
16 10 14.4	&	21 55 12	& 37.670	& 44.768	&	3.8	+/-	0.39	&	1.22	+/-	0.23	\\
16 10 19.2	&	21 52 48	& 37.624	& 44.739	&	2.4	+/-	0.46	&	1.05	+/-	0.28	\\
16 10 21.6	&	22 00 00 	& 37.789	& 44.765	&	3.1	+/-	0.53	&	0.99	+/-	0.26	\\
16 10 24.0	&	22 02 24	& 37.847	& 44.768	&	2.0	+/-	0.47	&	0.81	+/-	0.31\\	
16 10 26.4	&	21 57 36	& 37.743	& 44.735	&	2.1	+/-	0.53	&	0.81	+/-	0.32	\\
16 10 26.4	&	22 10 12	& 38.026	& 44.797	&	1.9	+/-	0.56	&	0.82	+/-	0.36\\	
16 10 31.2	&	21 40 12	& 37.361	& 44.632	&	1.6	+/-	0.54	&	1.53	+/-	0.40	\\
16 10 33.6	&	21 47 24	& 37.526	& 44.659	&	2.9	+/-	0.52	&	0.84	+/-	0.27	\\
16 10 33.6	&	21 55 12	& 37.700	& 44.697	&	1.9	+/-	0.44	&	0.83	+/-	0.31	\\
16 10 36.0	&	21 49 12	& 37.570	& 44.659	&	2.3	+/-	0.52	&	0.89	+/-	0.30	\\
16 10 38.4	&	21 45 00	& 37.480	& 44.629	&	4.1	+/-	0.57	&	1.44	+/-	0.24	\\
16 10 38.4	&	22 04 48	& 36.583	& 44.427 &	1.7	+/-	0.51	&	0.71	+/-	0.36	\\
16 10 40.8	&	21 52 48	& 37.658	& 44.659	&	1.7	+/-	0.34	&	0.70	+/-	0.28\\	
16 10 45.6	&	21 42 36	& 37.438	& 44.591	&	3.7	+/-	0.73	&	1.50	+/-	0.28	\\
16 10 52.8	&	21 40 12	& 37.396	& 44.552	&	2.9	+/-	0.68	&	1.62	+/-	0.31	\\
16 10 52.8	&	21 47 24	& 37.556	& 44.588	&	3.9	+/-	0.45	&	1.04	+/-	0.23	\\
16 10 57.6	&	21 37 12	& 37.336	& 44.520	&	3.5	+/-	0.79	&	3.28	+/-	0.31	\\
16 10 57.6	&	21 49 12	& 37.604	& 44.579	&	1.9	+/-	0.42	&	0.79	+/-	0.30\\	
16 11 00.0	&	21 45 00	& 37.514	& 44.549	&	3.5	+/-	0.21	&	0.85	+/-	0.21	\\
16 11 12.0	&	21 40 12	& 37.427	& 44.481	&	3.5	+/-	0.45	&	1.13	+/-	0.24	\\
16 11 19.2	&	21 37 12	& 37.371	& 44.440	&	2.9	+/-	0.45	&	1.78	+/-	0.26	\\
16 11 24.0	&	21 34 48	& 37.326	& 44.410	&	2.6	+/-	0.77	&	3.05	+/-	0.36	\\
16 11 24.0	&	21 55 12	& 37.781	& 44.511	&	1.2	+/-	0.37	&	1.27	+/-	0.37\\	
16 11 31.2	&	22 00 00	& 37.899	& 44.508	&	2.4	+/-	0.61	&	1.27	+/-	0.32\\	
16 11 43.2	&	21 55 12	& 37.811	& 44.440	&	2.0	+/-	0.48	&	1.69	+/-	0.31\\	
16 11 57.6	&	21 49 48	& 37.714	& 44.360	&	1.2	+/-	0.39	&	2.21	+/-	0.39\\	
\hline
\multicolumn{2}{l}{{$\mathrm{^{a}}$ Data from \citet{ccm10}.}}\\
\enddata
\end{deluxetable}
\clearpage

\begin{figure}
\includegraphics[width=150mm]{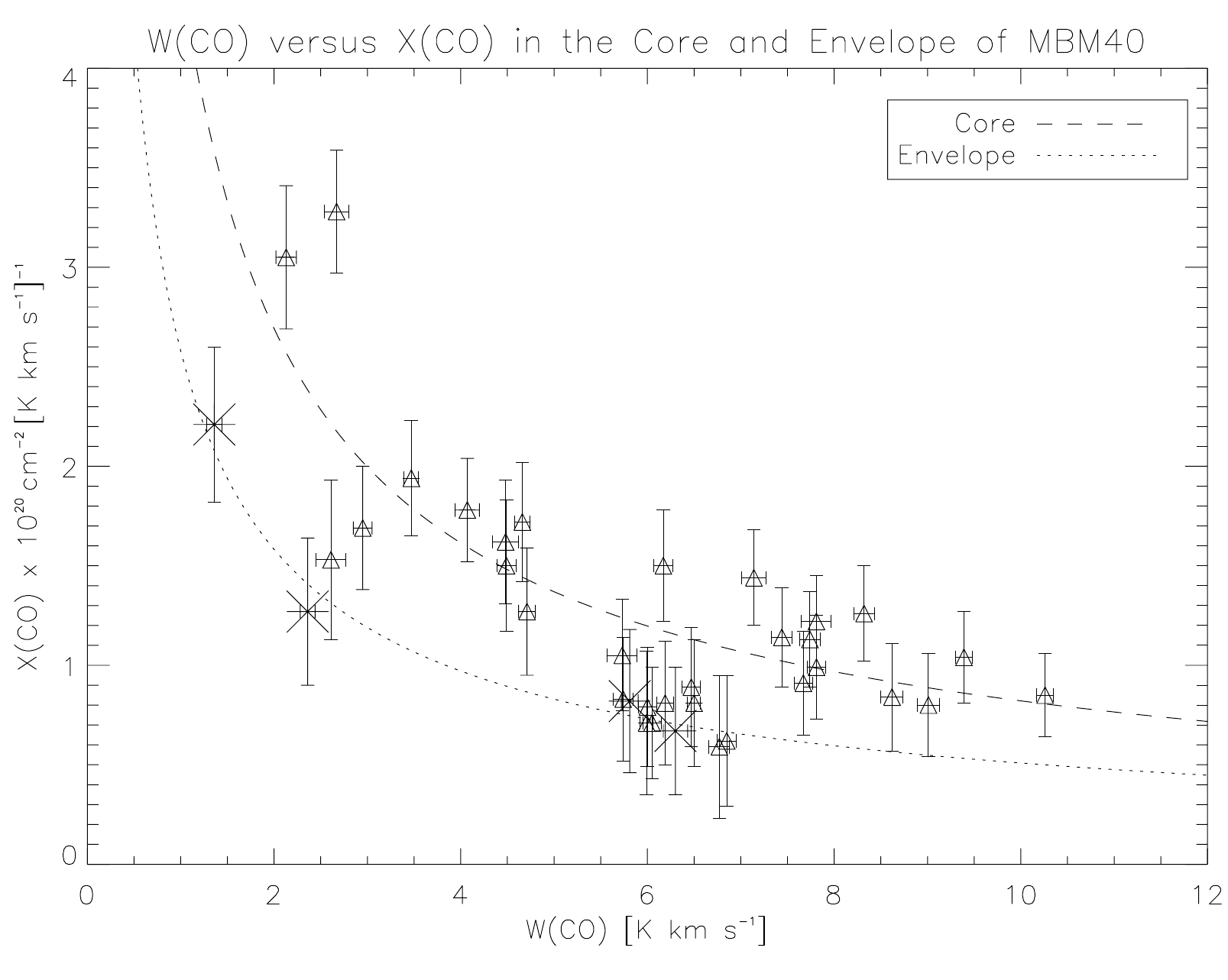}
\caption[$\mathrm {X_{CO}}$ versus $\mathrm{W_{CO}}$ for the envelope and the core of 
MBM 40]{$\mathrm {X_{CO}}$ versus $\mathrm{W_{CO}}$ for the envelope and the core of MBM 40. 
The envelope consists of four points and are labeled with 'X'. The curve fit for the core is 
of the form $\mathrm {X_{CO}=4.5\ x\ 10^{20}[W_{CO}^{-0.78}]}$ (with a  coefficient  
of determination of 0.54) and for the envelope it is of the form 
$\mathrm {X_{CO}=2.6\ x\ 10^{20}  [W_{CO}^{-0.71}]}$ (with a coefficient  of 
determination of 0.97).}\label{fig:xco-wco}
\end{figure}

Figure \ref{fig:xco-wco} shows how $\mathrm {X_{CO}}$ varies with respect to
$\mathrm{W_{CO}}$.  An inverse relationship seems to hold as first noted by
\citet{mag98}. It can be seen that in areas with high $\mathrm{W_{CO}}$ ($>$ 8 $\mathrm{K\
km\ s^{-1}}$), $\mathrm {X_{CO}}$ remains fairly constant at a level of $\sim$ 1
$\times$ 10$^{20}$.
 The other extreme of the graph ($\mathrm{W_{CO}}$ $\le$ 4 K km s$^{-1}$) is the section
with CO values 
 that most resembles those found in the envelope and periphery of MBM 40.  
The curve fit to the data shows that $\mathrm {X_{CO}}$ increases to 2-3 $\times$ 10$^{20}$. For
 the core region the curve fit follows the function $\mathrm {X_{CO}=4.5\ x\
 10^{20}[ \mathrm{W_{CO}}^{-0.78}]}$ with a  coefficient  of determination of 0.54,
 and for the envelope it is $\mathrm {X_{CO}=2.6\ x\ 10^{20}  [
 \mathrm{W_{CO}}^{-0.71}]}$ with a coefficient  of determination of 0.97.
By calibrating the value of X$_{CO}$ in these low-intensity CO(1-0) regions we
can determine N(H$_2$) more confidently than by using one global value for
all three cloud regions.

These trends are usually interpreted using simple models of photodissociation
 regions.   In the core of a dark cloud, virtually all the carbon is in the form
 of CO and so the H$_2$ and CO are well-correlated and a standard value for
 $\mathrm{X_{CO}}$ is reasonable.  In the outer envelope of a molecular cloud, away from
the dense core(s), the extinction is typical of the translucent regime and CO is
photo-dissociated more strongly than $\mathrm {H_{2}}$.  There, most
 carbon is in the form of C$\rm{I}$ and C$\rm{II}$ while the fraction of hydrogen
in molecular form is $\ge$ 50\% so that the $\mathrm{X_{CO}}$ value 
should increase as $\mathrm{W_{CO}}$
decreases \citep{vDB88}. Unfortunately, these models, which have been used
repeatedly to interpret the CO abundances and observations
 for molecular clouds with A$_V <$
than 5 mag do not encompass the full observational milieu. Recent
 work by \citet{lp12} establishes without any doubt that strong CO(1-0) 
emission (4-5 K or more in
 T$_{mb}$) can arise in regions with E(B-V) $\le$ 0.15 mag.
This is equivalent to an A$_V <$ 0.5 mag; no longer in the translucent regime
according to the definition of \citet{vDB88} but in the diffuse
molecular cloud regime.  Here, the CO/H$_2$ ratio and CO column densities 
should be too low for any detection of the CO(1-0) line, let alone a line greater
than 4-5 K in T$_{mb}$.  \citet{lp12} attribute these strong lines to radiative
transfer (sub-thermal excitation and scattering of photons in low-density
regions) and chemistry effects.  Given the macroscopic turbulence in these
low-density molecular clouds, it is clear the previous models of the CO photochemistry
were too simplistic and must be revised to account for a more realistic radiative
transfer model.

MBM 40, long identified as a translucent cloud, is instead more similar to the
diffuse molecular clouds with strong CO emission discussed by \citet{lp12}.
The peak E(B-V) in the direction of MBM 40 according the SFD dust maps is 0.24 mag equivalent
to an A$_V$ of 0.74 for the canonical value of R=3.1.  However,
even if the core of the cloud, defined as the wishbone-shaped region with
0.17 $\le$ E(B-V) $\le$ 0.24 mag, could be considered ``translucent" because of
an anomalously high value of R and/or a significantly lower than average interstellar radiation 
field, the envelope and periphery of the cloud are clearly diffuse molecular gas, and
some of the CO(1-0) lines in these regions exceed 5 K in T$_{mb}$.  MBM 40 is thus most 
likely a diffuse molecular cloud, even including the core region, 
rather than a translucent cloud as has been assumed
for the last 2 decades.

We can use the N(H$_{total}$) - E(B-V) relation established by \citet{BSD78}
to check our values for $\mathrm{X_{CO}}$ determined via the CH method.  Using
N(H$_{total}) = $ 5.8 $\times$ 10$^{21}$ [E(B-V)] yields a value of 
1.4 $\times$ 10$^{21}$ cm$^{-2}$ for the peak reddening position in MBM 40.
If we take a value of 7 $\times$ 10$^{19}$ cm$^{-2}$ for N(HI) in the direction 
of the peak \citep{shore03}, the derived value of N(H$_2$) based on the
reddening  is $\sim$ 7 $\times$ 10$^{20}$ 
cm$^{-2}$.   The value of $\mathrm{W_{CO}}$ closest to the peak reddening position yields
7.7 $\times$ 10$^{20}$ cm$^{-2}$ using a calibrated value of $\mathrm{X_{CO}}$
for the core region of 1 $\times$ 10$^{20}$ (see below).

\section{Mass of MBM 40} \subsection{Determination from $\mathrm{X_{CO}}$}

To determine the mass of $\mathrm{H_2}$ from our CO measurements we use the
$\mathrm{X_{CO}}$ values determined from Figure \ref{fig:xco-wco} in conjunction
with the observed, average $\mathrm{W_{CO}}$ values for each region.  In this study all
44 observations within the core region yielded detections, with an average
$\mathrm{W_{CO}}$ value of 5.02 $\pm$ 0.11 $\mathrm{K\ km\ s^{-1}}$.  In regions of lower
E(B-V), we detected CO in 28 of 38 locations in the envelope and 13 of 21 in
the periphery resulting in average $\mathrm{W_{CO}}$ values of 1.37 and 0.71
 $\mathrm{K\ km\ s^{-1}}$, respectively.  Guided by the inverse
relationship established by Figure \ref{fig:xco-wco}, a value of  1 x
$\mathrm{10^{20}}$ for $\mathrm {X_{CO}}$  will be used to calculate N(H$_2$) in
the core, while in the envelope and periphery, an $\mathrm {X_{CO}}$  value of
2 x $\mathrm{10^{20}}$ is deemed more appropriate.  We underscore that in MBM
40, the difference in E(B-V) between core and periphery is only a factor of 3 so
that even in the core a sizable fraction of the carbon may be in the form of
C$\rm{I}$. \citet{ing97}  found that in 10 high-latitude molecular clouds the
ratio of C/CO averaged 1.2 implying that translucent high-latitude molecular
clouds  are transitional objects between clouds dominated by CO and clouds where
most of the carbon is in atomic form.  As a diffuse molecular cloud, MBM 40 is likely
to have an even greater proportion of carbon in atomic form.

Our empirically-calibrated $\mathrm{X_{CO}}$ values vary by a factor of 2.  However, the
rise in $\mathrm{X_{CO}}$ as $\mathrm{W_{CO}}$ decreases is driven by only 7 data points
with $\mathrm{W_{CO}}$ $<$ 5 K km s$^{-1}$.  Thus, the uncertainty in determining the
mass of the various regions of the cloud is going to be large. The basic reason
for the lack of $\mathrm {X_{CO}}$ data in the envelope of the cloud is that
detecting CH emission from regions of very low E(B-V) is  difficult, requiring
several hours of integration per point.  In order to estimate the uncertainty in
our $\mathrm {X_{CO}}$ values, we can turn to the lower resolution CH data used
by \citet{mag98} to calibrate $\mathrm {X_{CO}}$ in the outer regions of MBM 40.  These data
were at 9$\arcmin$ and 8$\arcmin$ resolution for the CH and CO transitions,
respectively, but 7 of the 11 data points from that study were in the envelope
and periphery of the cloud.  The value of $\mathrm {X_{CO}}$ obtained from that
data is 3.15 x $\mathrm{10^{20}}$.  Though at lower resolution, this value can
be used as and independent measure of $\mathrm {X_{CO}}$ outside the core of the
cloud.  Thus, we will use 2 $\pm$ 1 x $\mathrm{10^{20}}$ as our estimate of the
value of  $\mathrm {X_{CO}}$.  This uncertainty (which drives the overall uncertainty in
the mass of the cloud for the envelope and periphery) will thus be estimated at $\sim$ 50\%

\begin{deluxetable}{ l   c c c c c}
\tablecaption{Average values for the detections in MBM 40.\label{tab:avgco}}
\tablewidth{0pt}
\tablehead{
\colhead{Region} & \colhead{$\mathrm{T_{R}\ ^{*}}$} & \colhead{$\mathrm {\Delta v}$(FWHM)}  &
\colhead{$\mathrm {v_{LSR}}$} & \colhead{$\mathrm {\sigma_{LSR}}$}    &  \colhead{$\mathrm {W_{CO}}$ $\mathrm{^{a}}$}\\
& \colhead{(K)} & \colhead{(km s$^{-1}$)} & \colhead{(km s$^{-1}$)} & \colhead{(km s$^{-1}$)}   
&  \colhead{(K km s$^{-1}$)}
}
\startdata
Core 		&	4.57 +/- 2.40	&	0.90 +/- 0.20 & 	3.20 & 0.25 	&	5.02 +/- 0.11    	\\
Envelope 		&  	1.43 +/- 1.51	&	0.71 +/- 0.24 & 	3.19 & 0.29	&	1.37 +/- 0.08		\\
Periphery 		&	0.78 +/- 0.68	&	0.76 +/- 0.17 & 	3.51 & 0.66	&	0.71 +/- 0.06		\\
All points		&	2.96 +/- 2.57	&	0.82 +/- 0.23 	&	3.25 & 0.37 	& 	3.16 +/- 0.09		\\
\hline
\multicolumn{2}{c}{{$\mathrm{^{a}}$} $\mathrm{W_{CO}}$ is $\mathrm{\int T_{mb}\ dv}$.}
\enddata
\end{deluxetable}

To obtain the mass of MBM 40 in terms of H$_2$ using CO observations the 
regions where CO is measured must be extrapolated over the entire of the cloud. 
Using the average $\mathrm{W_{CO}}$ values from Table \ref{tab:avgco} with the 
aforementioned  $\mathrm {X_{CO}}$ values of 2 and 1 x $\mathrm{10^{20}}$ for 
the two outermost regions and the core, respectively, the column density of molecular hydrogen can be calculated.
 The distance of the cloud is usually estimated to be 130 $\pm$ 10 pc \citep[see, e.g.][]{ccm10}. 
Combining the average N(H$_2$) for each region,  the ratio of detections/observations for each region 
($\beta$), the angular size of each region in steradians ($\Omega$), the distance to the cloud (d),
and  the mean molecular mass, $\mu$ (taken to
be 2.3 to account for helium) allows for a determination
of the mass: 
\begin{equation}
\label{eq:mass-h2}
\mathrm{M_{H_2} = \beta\ \Omega\ d^{2}\  N(H_{2})\ \ \mu\  m_H \ \ \ \ (gm)} 
\end{equation}

Using this method yields molecular masses of  9.6 $\pm$ 5, 12 $\pm$ 6, and  
10 $\pm$ 5 M$_{\sun}$ for the core, envelope, and periphery regions, respectively. 
These masses are about
a factor of 2 greater than those determined from OH observations by \citet{cotten12a} who estimated molecular 
masses of 3.8 $\pm$ 0.86, 7.6 $\pm$ 2.5, and 5.2 $\pm$ 6.3 M$_{\sun}$ for  the core, 
envelope, and periphery regions, respectively.   The differences in the two estimates depend
directly on the values of $\mathrm {X_{CO}}$   and the OH/H$_2$ abundance that are used
in each derivation of the mass for each region.  Given the uncertainties in both quantities, the agreement is satisfactory.
Both the CO and OH data show that MBM 40 is a small, diffuse molecular cloud, with a mass in the 15 - 30 M$_\odot$ range.
More importantly, like \citet{cotten12a}, we find that as much as 
a third of the total cloud mass may be found in the periphery where E(B-V) $<$ 0.12 mag.
This is significant because molecular cloud mapping in CO seldom probes regions of
such low extinction.  If substantial molecular gas is found here, then at least some of the
``dark" molecular gas found in recent studies \citep{grenier05} may  be spectroscopically detectable
by radio means after all.

\subsection{Virial Considerations}

In addition to determining the mass of the cloud via an $\mathrm {X_{CO}}$
value, our CO observations allow us to determine the virial mass of the cloud
and thus determine the gravitational state of MBM 40.  Assuming a density
distribution that scales as $\rho^{-2}$ from the core of the cloud, a cloud in virial equilibrium should
have a virial mass, M$_{vir}$, in solar masses of 126 $r$ $\Delta v^2_{tot}$, where
$\Delta v_{tot}$ is the velocity width (FWHM) of the composite spectrum over the entire cloud,
and $r$ is the radius of the cloud in parsecs.
We can estimate $\Delta v_{tot}$ following \citet{dick85} who
determine the velocity dispersion of the composite spectral profile of the 
cloud ($\sigma_p$) from the dispersion of the average velocity width for the entire cloud
($\sigma_i$), and the dispersion of the centroid velocities for the whole cloud ($\sigma_c$):

$$ \sigma_p^2 = \sigma_i^2 + \sigma_c^2 $$

\noindent
In our case, we can use the data from Table 2, where $\sigma_c$ is equivalent
to $\sigma_{LSR}$ and $\sigma_i$ is just the dispersion of the line widths 
averaged for the whole cloud.   In this manner, we obtain $\sigma_p =$ 
0.51 km s$^{-1}$ or a $\Delta v_{tot} =$ 1.2 km s$^{-1}$.  Using a distance of 130 $\pm$ 10 pc, the radius of the cloud can be taken to be
the (A/$\pi$)$^{0.5}$ assuming the cloud is spherical in shape (a reasonable 
assumption - see Figure 1).  At 130 pc, this gives a radius of 0.8$^\circ$, equivalent to 1.8 pc.
With these values, M$_{vir} =$ 330 M$_\odot$.  
Even including a contribution to the mass from the HI associated with the cloud (estimated
by \citet{RJC05} to be 10 M$_\odot$), we see that cloud is clearly
gravitationally unbound - a typical result for a high-latitude cloud and 
in keeping with its categorization as a diffuse molecular cloud.


\section{Summary}

We observed 103 lines of sight in the high-latitude cloud, MBM 40, in the CO(1-0) transition
using the Arizona Radio Observatory 12 m radio telescope, detecting emission from 85
positions.  The cloud was divided into 3 regions based on the reddening maps of SFD:
a core region where 0.25 $>$ E(B-V) $>$ 0.17 mag, an envelope region where
0.12 $\le$ E(B-V) $\le$ 0.17, and an outermost periphery region where E(B-V)  $<$
0.12 mag.  The detection rates for each region were 44/44 in the core, 28/38 in the 
envelope, and 13/21 in the periphery.

Using previously published CH data, we calibrated the value of the CO-H$_2$ conversion
factor, $\mathrm{X_{CO}}$, for this cloud.  The values we obtained, ranging from 0.6 $\times$ 10$^{20}$
 to 3.3 $\times$ 10$^{20}$ cm$^{-2}$ 
[K km s$^{-1}$]$^{-1}$ with an average of 1.3  $\times$ 10$^{20}$ cm$^{-2}$ 
[K km s$^{-1}$]$^{-1}$, are similar to what was obtained at nearly an order of magnitude
worse resolution by \citet{mag98}.  Like those authors, we find an inverse relationship
between $\mathrm{X_{CO}}$ and $\mathrm{W_{CO}}$.  This has also been noted by other authors in diffuse
molecular clouds \citep[e.g.][]{lpl10}.

This cloud has a peak reddening of 0.24 mag which for a normal value of R places it
squarely in the category of diffuse molecular clouds following the schema of \citet{vDB86}.
 This is in contrast to nearly all previous papers on this object, the authors
of which categorized this object as a translucent molecular cloud.  It is surprising that 
a diffuse molecular cloud should have CO(1-0) emission as intense as that seen in this
object.  Recently, \citet{lp12}. have made a convincing case for a class of diffuse
molecular cloud with strong CO(1-0) lines, similar to those from denser, more opaque
molecular clouds.  The intense CO emission is attributed to radiative transfer and
chemistry effects.  We believe this to be also the case for MBM 40.

With our calibrated values of $\mathrm{X_{CO}}$ for the core and outer regions of the cloud, we
can determine the cloud mass in each region and overall.  The values we obtain:
9.6, 12, and 10 M$_\odot$ for the core, envelope, and periphery, respectively, are similar to what was found
by \citet{cotten12a} using the OH 1667 MHz line as a molecular tracer.    
A virial analysis shows that MBM 40 is not gravitationally bound. Both studies show that as much as 1/3 of the cloud's molecular mass may be in the
outermost regions of the cloud
where the visual extinction is likely to be no more than 0.4-0.5 magnitudes.

In summary, MBM 40 is a small, nearby, diffuse molecular cloud with strong CO(1-0)
emission.  Like other diffuse molecular clouds, it is not gravitationally bound and 
destined to break up over the sound-crossing time scale (of order 10$^6$ years).
As such, the object is not a candidate for star formation as has been confirmed by
several studies \citep{minh03, shore03, mag96}.


We would like to thank Lucy Ziurys and the University of Arizona for providing
support for this project, and to an anonymous referee for many useful suggestions.


\bibliographystyle{aa} 
\bibliography{CO_Final}






\end{document}